\documentstyle[11pt,newpasp,twoside,epsf]{article}
\def\edcomment#1{\iffalse\marginpar{\raggedright\sl#1\/}\else\relax\fi}
\reversemarginpar

\begin{document}
\title{Microquasars as sources of high energy phenomena}

\author{I.F. Mirabel}
\affil{Centre d'Etudes de Saclay/ CEA/DSM/DAPNIA/SAP\\
           91911 Gif/Yvette, France \& \\
           Intituto de Astronom\'\i a y F\'\i sica del Espacio/CONICET, Argentina}

\begin{abstract} 
Relativistic outflows are a common phenomenon in accreting black
holes. Despite the enormous differences in scale, accreting 
stellar-mass black
holes (X-ray binaries, collapsars) and super-massive black holes produce jets with analogous physical properties. Here I review microquasars as  
sources of relativistic jets, gamma-rays, cosmic rays, and high energy neutrinos.
\end{abstract}

\section{The quasar-microquasar analogy}

Microquasars are scaled-down versions of quasars and both are believed to be
powered by spinning black holes with masses of up to a few tens that
of the Sun (see Figure 1). The word {\it microquasar} was chosen by 
Mirabel et al. (1992) to suggest that we could learn about microquasars from 
previous decades of studies on AGN. 
In fact, the analogy with quasars is more than morphological, because 
there is an underlying unity in the physics of accreting black
holes over an enormous range of scales, from stellar-mass black holes
in binary stellar systems, to super-massive black holes at the centre
of distant galaxies (Rees, 1998). A major difference is that 
the linear and time scales of the phenomena are proportional to the black 
hole mass. 

In quasars and microquasars are found
the following three basic ingredients: 1) a spinning black hole, 2) an
accretion disk heated by viscous dissipation, and 3) collimated jets
of relativistic particles. However, in microquasars the black hole is
only a few solar masses instead of several million solar masses; the
accretion disk has mean thermal temperatures of several million degrees
instead of several thousand degrees; and the particles ejected at
relativistic speeds can travel up to distances of a few light-years
only, instead of the several million light-years as in some giant radio
galaxies. In quasars matter can be drawn into the accretion disk from
disrupted stars or from the interstellar medium of the host galaxy,
whereas in microquasars the material is being drawn from the companion
star in the binary system. In quasars the accretion disk has sizes of
$\sim$10$^9$ km and radiates mostly in the ultraviolet and optical
wavelengths, whereas in microquasars the accretion disk has sizes of
$\sim$10$^3$ km and the bulk of the radiated energy comes out in the
X-rays. It is believed that part of the spin energy of the black hole
can be tapped to power the collimated ejection of magnetized plasma at
relativistic speeds.  

Because of the
relative proximity and shorter time scales, in microquasars it is
possible to firmly establish the relativistic motion of the sources of
radiation, and to better study the physics of accretion flows and jet
formation near the horizon of black holes. Jets in microquasars are easier 
to follow because their apparent motions in the sky are $\geq$10$^3$ 
faster than in quasars. Because microquasars are found in our Galaxy 
the two-sided moving jets are more easily seen than in AGN (Mirabel \& 
Rodr\'\i guez, 1994). However, 
to know how the 
jets are collimated in units of length of the black hole's horizon, AGN up 
to distances of a few Mpc may present an advantage. Biretta et al. (2002) 
find that the initial collimation of the non-thermal jet in the galaxy 
M87 of the Virgo cluster takes place on a scale of 30-100 R${_S}$, which 
is consistent with poloidal collimation by an accretion disk. 

At first glance it may seem paradoxical that relativistic jets were
first discovered in the nuclei of galaxies and distant quasars and
that for more than a decade SS433 was the only known object of its
class in our Galaxy (Margon 1984). This is because that disks
around super-massive black holes emit strongly at optical and UV
wavelengths.  Indeed, the more massive the black hole, the cooler the
surrounding accretion disk is.  For a black hole accreting at the
Eddington limit, the characteristic black body temperature at the last
stable orbit in the surrounding accretion disk will be given
approximately by $T \sim 2 \times 10^7~M^{-1/4}$ (Rees 1984), with $T$
in K and the mass of the black hole, $M$, in solar masses.  Then,
while accretion disks in AGN have strong emission in the optical and
ultraviolet with distinct broad emission lines, black hole and neutron
star binaries usually are identified for the first time by their X-ray
emission. 
Among this class of sources, SS433 is unusual given its broad
optical emission lines and its brightness in the visible. Therefore,
it is understandable that there was an impasse in the discovery of new
stellar sources of relativistic jets until the development of
X-ray astronomy that started with the discovery of the first extra-solar 
X-ray source by Giacconi et al. (1962). Strictly speaking, if it had 
not been for the historical circumstances described above, the 
acronym {\it quasar}
would have suited better the stellar mass versions rather than their
super-massive analogs at the centers of galaxies.

\begin{figure}[h]
\plotone{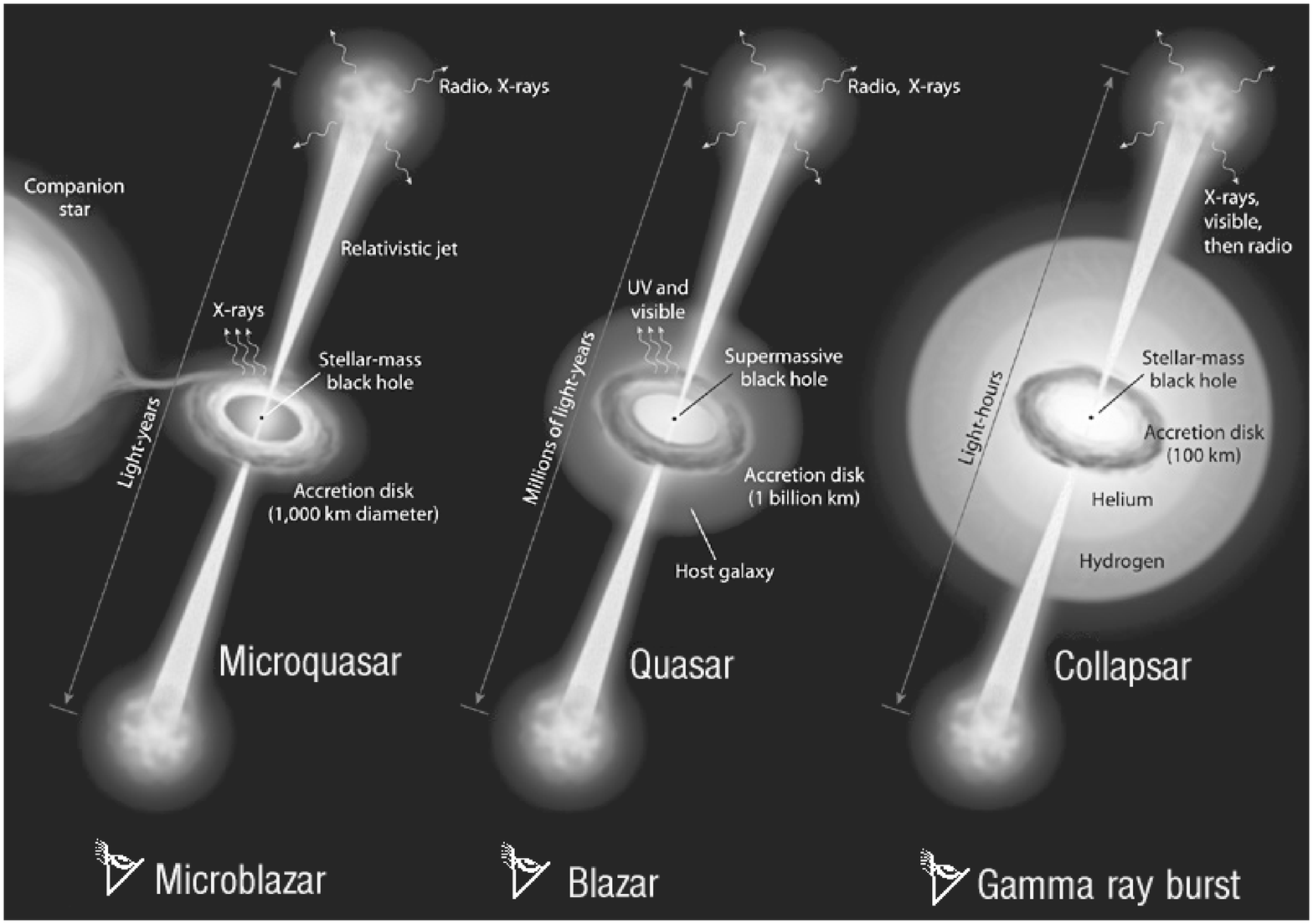}
\caption{Diagram illustrating current ideas concerning microquasars, quasars 
and gamma-ray bursts (not to scale). It is proposed that a universal 
mechanism may be at work in all sources of relativistic jets in the 
universe. Synergism between these three areas of research in astrophysics 
should help to gain a more comprehensive understanding of the relativistic 
jet phenomena observed everywhere in the universe.} 
\label{fig-1}
\end{figure}

\section{The microquasar gamma-ray-burst analogy}

There is increasing evidence that the central engine of the most
common form of gamma-ray burst (GRBs), those that last longer than a
few seconds, are afterglows from ultra-relativistic jets produced
during the formation of stellar-mass black holes (McFaden \& Woosley,
1999). Mirabel \& Rodr\'\i guez (1999) proposed that ultra-relativistic
bulk motion and beaming are needed to explain: 1) the enormous energy
requirements of $\geq$ 10$^{54}$ erg if the emission were isotropic
(e.g. Kulkarni et al. 1999; Castro-Tirado et al. 1999); 2) the
statistical correlation between time variability and brightness
(Ramirez-Ruiz \& Fenimore, 2000), and
3) the statistical anti-correlation between brightness and time-lag
between hard and soft components (Norris et al. 1999). Beaming reduces
the energy release by the beaming factor f = $\Delta$$\Omega$/4$\pi$,
where $\Delta$$\Omega$ is the solid angle of the beamed
emission. Additionally, the photon energies can be boosted to higher
values.  Extreme flows from collapsars with bulk Lorentz factors $>$
100 have been proposed as sources of $\gamma$-ray bursts (M\'esz\'aros
\& Rees 1997). High collimation (Dado, Dar \& de R\'ujula 2002; Pugliese et al. 1999) can
be tested observationally (Rhoads, 1997), since the statistical
properties of the bursts will depend on the viewing angle relative to
the jet axis.

Recent multi-wavelength studies of gamma-ray afterglows suggest that
they are highly collimated jets. The brightness of the optical
transient associated to some GRBs show a break (e.g. Kulkarni et
al. 1999), and a steepening from a power law in time t proportional to
t$^{-1.2}$, ultimately approaching a slope t$^{-2.5}$ (e.g. Castro-Tirado
et al. 1999). The achromatic steepening of the optical light curve and
early radio flux decay of some GRBs are inconsistent with simple
spherical expansion, and well fit by jet evolution. It is interesting
that the power laws that describe the light curves of the ejecta in
microquasars show similar breaks and steepening of the radio flux
density (Rodr\'\i guez \& Mirabel, 1999). In microquasars, these
breaks and steepenings have been interpreted (Hjellming \& Johnston
1988) as a transition from slow intrinsic expansion followed by free
expansion in two dimensions. Besides, linear polarizations of about
2\% were recently measured in the optical afterglows 
(e.g. Covino et al. 1999), providing strong evidence that the afterglow
radiation from gamma-ray bursts is, at least in part, produced by
synchrotron processes.  Linear polarizations in the range of 2-10\%
have been measured in microquasars at radio (e.g. Rodr\'\i guez et
al. 1995), and optical (Scaltriti et al. 1997) wavelengths.

The jets in microquasars of our own Galaxy seem to be
less extreme local analogs of the super-relativistic jets associated
to the more distant gamma-ray bursts. But the latter do not repeat, 
seem to be
related to catastrophic events, and have much larger super-Eddington
luminosities. According to the latest models, the same symbiotic disk-jet 
relationship as in microquasars and quasars powers the GRBs. In fact, 
it is now believed that the Lorentz factors at the base of the jets inside the 
collapsing star are $\leq$10 as in microquasars and quasars, 
and they reach values $\geq$100 
when they break free from the infalling outer layers of the 
progenitor star. Because of the enormous difference in power, 
the scaling laws in terms of the black hole
mass that are valid for the analogy between microquasars and quasars  
may not apply in the case of gamma-ray bursts.

\section{Compact jets in stellar-mass and super-massive black holes}

The class of stellar-mass black holes that are persistent X-ray
sources (e.g. Cygnus X-1, 1E 1740-2942, GRS 1758-258, etc.) and some
super-massive black holes at the centre of galaxies (e.g. Sgr A$^*$ and
many AGN) do not exhibit luminous outbursts with large-scale sporadic
ejections. However, despite the enormous differences in mass, steadily
accreting black holes have analogous radio cores with steady, flat
(S$_{\nu}$$\propto$$\nu$$^{\alpha}$; $\alpha$$\sim$0) emission at
radio wavelengths. The fluxes of the core component in AGN are
typically of a few Janskys (e.g. Sgr A$^*$$\sim$1Jy) allowing VLBI
high resolution studies, but in stellar mass black holes the cores are
much fainter, typically of a few mJy, which makes difficult
high resolution observations of the core.

From the spectral shape it was proposed that 
the steady compact radio emission in black hole X-ray binaries are jets
(e.g. Rodr\'\i guez et al. 1995; 
Fender et al. 1999, 2000; Corbel et al. 2000). 
Recently, this has been confirmed by VLBI observations 
at AU scale resolution of GRS 1915+105 (Dhawan, Mirabel \&
Rodr\'\i guez, 2000), and Cyg X-1 (Stirling et al. 2001) in the low-hard 
X-ray state. VLBA images of GRS 1915+105 show compact jets with sizes $\sim$10$\lambda$$_{cm}$ AU along the
same position angle as the superluminal large-scale jets. 
As in the radio cores of AGN, the brightness temperature of the
compact jet in GRS 1915+105 is T$_B$$\geq$10$^9$ K. The VLBA images of
GRS 1915+105 are consistent with the conventional model of a conical
expanding jet with synchrotron emission (Hjellming \& Johnston, 1988;
Falcke \& Biermann, 1999) in an optically thick region of solar system
size. These compact jets are also found in neutron star X-ray binaries 
such as LS 5039 (Paredes et al. 2000) and Sco X-1 (Fomalont et al. 2001),  
and are currently used to track the  
path of black holes and neutron stars in our Galaxy 
(see Mirabel \& Rodrigues, 2002 for a review). 

\begin{figure}
\plotone{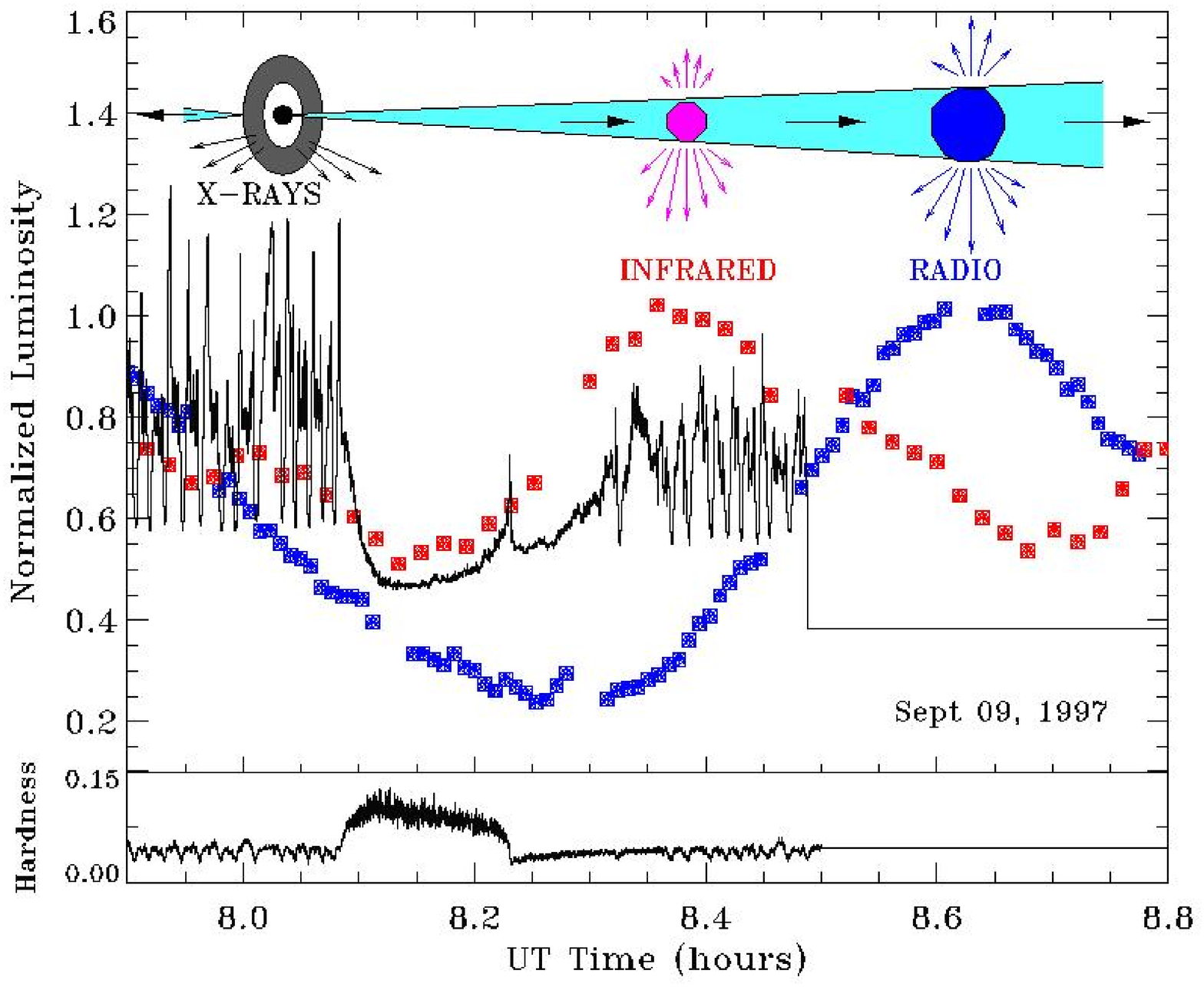}
\caption{Radio, infrared, and X-ray light curves
for GRS~1915+105 at the time of quasi-periodic oscillations with  
scales of time of $\sim$20 min (Mirabel et al. 1998).  
The infrared flare starts during
the recovery from the X-ray dip, when a sharp, isolated X-ray spike-like 
feature  is
observed.  These observations show the connection between the rapid
disappearance and follow-up replenishment of the inner accretion disk
seen in the X-rays (Belloni et al. 1997), with the ejection of
relativistic plasma clouds observed  first as synchrotron emission at
infrared wavelengths, later at radio wavelengths.  A scheme
of the relative positions where the different emissions originate is
shown in the top part of the figure.  The hardness ratio (13-60
keV)/(2-13 keV) is shown at the bottom of the figure. 
Analogous phenomena have now been obseved  
in the quasar 3C 120 but in time scales of years (Marscher et al. 2002).} \label{fig-2}
\end{figure}

\section{Accretion disk origin of relativistic jets}

Synergism between results from multiwavelength simultaneous observations 
in microquasars and quasars is providing important insights into the 
connection between accretion disk instabilities and the genesis of jets. 
Since the characteristic times in the flow of matter onto a black hole
are proportional to its mass, the accretion-ejection phenomena in 
quasars should last 10$^5$-10$^7$ longer than analogous phenomena 
in microquasars (Sams et al. 1996).  Therefore,
variations on scales of tens of minutes of duration in microquasars 
could be sampling
phenomena that had been difficult to observe in quasars. 

Simultaneous
multiwavelength observations of a microquasar revealed in an interval of 
time of a few tens of minutes the connection between the sudden 
disappearance of the inner $\sim$200 km of the accretion disk 
with the ejection of expanding clouds of relativistic plasma (see Figure 2). 
One possible interpretation of the observations shown in Figure 2 is that the plasma 
of the inner disk that radiates in the X-rays falls beyond the horizon 
of the black hole in $\sim$5min, and subsequently the inner accretion disk is 
refilled in $\sim$20 min. While the inner disk is being replenished, 
we observe the ejection of a relativistic plasma cloud, first 
at 2$\mu$m, and latter at radio wavelengths as the cloud expands and becomes 
transparent for its proper radiation at longer wavelengths. 
The delay between the maxima at radio and 
infrared wavelengths is equal to the one computed with the model for 
a spherically symmetric expanding clouds in relativistic AGN jets by 
van der Laan (1966). 
Although VLBA images of these transient ejecta by Dhawan et al. (2000) 
have shown that they are in fact connical jets, the model 
first developed for AGN is a good first approximation, and allows  
to demonstrate 
that the infared flares that preceed the radio flares are synchrotron,  
rather than thermal emission. This implies the presence in the jets of 
electrons with Lorentz factors $\geq$ 10$^3$ (Fender \& Pooley, 1998; Mirabel 
et al. 1998). 

Analogous accretion disk-jet connections were observed in the quasar 3C 120 
by Marscher et al. (2002). Jets were detected with VLBI after 
sudden X-ray dips observed with RXTE, but on scales of a few years. 
The scales of time of the 
phenomena are within a factor of 10 the black hole mass ratios between 
the quasar and microquasar, which is relativelly small when 
compared with the uncertainties in the data.

\section{Microquasars as sources of cosmic rays}

If a compact source injects 
relativistic plasma into its environment, it is expected
that some fraction of the injected power will be dissipated by shocks, 
where reacceleration of particles may take place. Evidences of 
such interactions are 
the radio lobes of 1E 1740.7-2942 (Mirabel et al. 1992), 
GRS 1758-258 (Rodr\'\i guez et al. 1992),  and the two lateral 
extensions in the nebula W50 that hosts at
its center SS 433. The interaction of SS 433 with the shells of  
W50 has been studied in the
X-rays (Brinkmann et al. 1996) 
and radio wavelengths (Dubner et al. 1998 and references therein).

Besides the well known
relativistic jets seen at sub-arcsec scales in the radio, large-scale
jets become visible in the X-rays
at distances $\sim$ 30 arcmin ($\sim$ 25 pc) from the
compact source (Brinkmann et al. 1996). In the radio and X-rays,
the lobes reach distances of up to 1$^\circ$ ($\sim$50 pc).
These large-scale X-ray jets and radio lobes
are the result of the interaction of the mass outflow with the
interstellar medium. From optical and X-ray emission lines it is found that
the sub-arcsec relativistic jets have a kinetic energy of $\sim$ 10$^{39}$ erg s$^{-1}$
(Margon, 1984), which is several orders of magnitude larger than the energy
radiated in the X-rays and in the radio.  Dubner et al. (1998)
estimate that the kinetic energy transferred into the ambient medium is
$\sim$ 2 10$^{51}$ ergs, thus confirming that the relativistic jets from SS433
represent an important contribution to the overall energy budget of 
the surrounding nebula W50. 

Large-scale, decelerating relativistic jets from the microquasar 
XTE J1550-564 have been discovered with CHANDRA by Corbel et al. (2002). 
The broadband spectrum of the jets is consistent with synchrotron 
emission from electrons with Lorentz factors of $\sim$10$^7$ that are 
probably accelerated 
in the shock waves formed by the interaction of the jets with 
the interstellar medium. Corbel et al (2002) demonstrated that in 
microquasars we can study in real time the formation and dynamical evolution of 
the working surfaces (lobes) of relativistic jets far away from the 
centres of ejection, on time scales inaccessible for AGN. Working 
surfaces of microquasar jets as in SS433 and XTE J1550-564 are potential 
sources of cosmic rays. 

In fact, Heinz \& Sunyaev (2002) propose that if microquasar jets contain 
cold protons and 
heavy ions (as in the case of SS433) thay can produce a small but measurable 
contribution to the cosmic ray spectrum in the range of 3-10 GeV. 
These authors propose that if the cosmic ray microquasar proton component 
is ruled out by observations, one could put interesting constraints 
on the particle content in microquasar jets by the observation of 
the 511 keV line with INTEGRAL. The process 
of cosmic ray production in microquasar jets is fundamentally different from 
the cosmic ray production in the non relativistic shocks of supernovae 
remnants. 
 
\section{Microquasars as sources of high energy neutrinos}

If the energy content of the jets in transient sources is 
dominated by electron-proton plasma as in SS433, 
Levinson and Waxman (2001) predict that an outburst of several hours 
of 1-100 TeV neutrinos should preceed the radio and infrared flares 
associated with major ejection events as the one observed by 
Mirabel \& Ridr\'\i guez (1994) in GRS 1915+105. These high energy 
neutrinos are produced by the interaction
of the high energy protons with synchrotron photons emitted by the
shock-accelerated electron. These neutrinos would provide 
further probe of the microquasar jet physics. 

Guetta et al. (2002)
estimate the neutrino fluxes produced in the jets of a sample of 
identified microquasars and microquasar candidates. They demonstrate 
that in several of the sources considered, the neutrino flux at Earth,
produced in events similar to those observed, can exceed the detection 
threshold of a km$^2$ neutrino
detector.  Neutrino and gamma-ray emission may be the way to identify 
microquasars candidates with jets directed along our line of sight.

\section{Microquasars as transient and persistent sources of gamma-rays}

Microblazars should exist and be 
found with the developement of new tools in astronomy. Mirabel \& 
Rodr\'\i guez (1999) proposed that for objects with angles between the 
line of sight and the jet axis $\theta \leq 10^\circ$ one expects the
timescales to be shortened by 2$\gamma$$^2$ and the flux densities to
be boosted by 8$\gamma^3$ with respect to the values in the rest frame
of the condensation.  For instance, for motions with $v$ = 0.98c
($\gamma$ = 5), the time-scale will shorten by a factor of $\sim$50 and
the flux densities will be boosted by a factor of $\sim 10^3$. Then,
for a galactic source with relativistic jets and small $\theta$ we
expected fast and intense variations in the observed flux.  
Microblazars are hard to detect in practice, both because of
the low probability of small $\theta$ values and because of the fast
decline in the flux. 

Recently, several flares were detected  with Konus from a region that 
contains Cygnus X-1
(Golenetskii et al. 2002), later confirmed with the BATSE data base 
as gamma-ray transients (Schmidt, 2002). These events have been interpreted 
by Romero, Kaufman Bernad\'o \& Mirabel (2002) in the context of the 
ideas  of Georganopoulos, Aharonian, \& Kirk (2002), as non-thermal emission by 
inverse Compton interaction between relativistic electrons in a precessing jet 
and external photon fields, with a dominant contribution from the companion 
star field  (Kaufman Bernad\'o, Romero \& Mirabel, 2002). Another microblazar 
could be V4641 Sgr, which exhibits outbursts with rapid optical 
variations (Uemura et al. 2002) and seems to 
be a source of jets with apparent speeds 
$\geq$10 (Orosz et al. 2001).

Paredes et al. (2000) found that the runaway persistent microquasar 
LS 5039 (Rib\'o et al. 2002) is inside the error box of the 
unidentified Egret source 3EG J1824-1514. In addition, a TeV source   
inside the core of the
OB association Cygnus OB2 and at the edge of the 95\% error circle of 
the EGRET source 3EG J2033+4118 has been reported by Aharonian et al (2002).  
No counterpart for the TeV source at other wavelengths has so far been 
identified, and one could envisage gamma -ray production via a jet-driven termination shock. Although most may be isolated neutron 
stars, a microquasar sub-population of EGRET sources is at this time 
a working hypothesis. 

I point out that Cygnus X-1, V4641 Sgr, and LS 5039 are all three confirmed 
sources of relativistic jets in high mass X-ray binaries. In this class of 
sources the jets 
initially move inside the UV photon field of the massive 
donor stars and inverse Compton of the jets with the UV 
photons is unavoidable. 

\vskip .1in

INTEGRAL and other space and ground based observatories for gamma-rays as 
well as km$^2$  
neutrino telescopes will in the next decade enhance the number of known microquasars 
providing unprecedented insights into the physics of high energy processes and 
phenomena in accreting stellar-mass black holes.

\end{document}